\documentclass[onecolumn]{revtex4}
\usepackage{graphicx}
\usepackage{amssymb}

\def\cO{{\mathcal O}}

\def\11{{\mathbb 1}}

\def\GeV{\,\rm GeV}

\def\cd{\!\cdot\!}

\def\beq{\begin{equation}}
\def\eeq{\end{equation}}
\def\bea{\begin{eqnarray}}
\def\eea{\end{eqnarray}}

\begin{document}
\title{A stable supermassive charged gravitino?}
\author{Krzysztof A. Meissner$^1$ and Hermann Nicolai$^2$\\}
\affiliation{\\
$^1$Faculty of Physics,
University of Warsaw\\
Pasteura 5, 02-093 Warsaw, Poland\\
$^2$Max-Planck-Institut f\"ur Gravitationsphysik
(Albert-Einstein-Institut)\\
M\"uhlenberg 1, D-14476 Potsdam, Germany
}

\vspace{10mm}

\begin{abstract} Some time ago it was suggested that dark matter may consist 
in part of an extremely dilute gas of supermassive gravitinos with fractional
charge 2$e$/3 \cite{MeissnerNicolai2019}. This scheme makes the definite (and 
falsifiable) prediction that massive gravitinos are the {\em only} new fermionic 
degrees of freedom beyond the known three generations of quarks and leptons 
of the Standard Model of Particle Physics. In this note we re-examine one 
special outlier event reported and subsequently discarded by the MACRO 
collaboration \cite{MACRO1} in the light of this proposal and 
point out the possibility of an alternative interpretation of this event
supporting the above hypothesis, whose confirmation (or refutation) would, 
however, require an independent dedicated experimental effort.
\end{abstract}
\maketitle

\vspace{5mm}

\section{Introduction.} 
The nature of dark matter (DM) continues to be one of the most vexing
questions of modern physics. While current DM scenarios are usually based 
on the assumption of ultralight constituents (such as axions) or 
TeV scale WIMPs, the possibility has been raised in recent work \cite{MeissnerNicolai2019}
that DM could consist at least in part of an extremely dilute gas  of 
supermassive stable gravitinos with charge $q=\pm\frac23$ in units of the
elementary charge $e$. This proposal has its roots in Gell-Mann's old observation 
that the fermion content of the Standard Model of Particle Physics (SM)
with three, and only three, generations of quarks and leptons (including
right-chiral neutrinos) can be matched with the spin-$\frac12$ content of the
maximal $N\!=\!8$ supermultiplet after removal of eight Goldstinos \cite{GM,NW}.
The only extra fermions beyond the known three generations of SM fermions would 
thus be eight massive gravitinos, but nothing else.
Crucially, the matching of U(1)$_{em}$ charges requires 
a `spurion shift' of $\delta q = \pm \frac16$ \cite{GM} that is not part of $N\!=\!8$ 
supergravity and that, in terms of the original spin-$\frac12$
fermions of $N\!=\!8$ supergravity, takes the very special form given in \cite{MN2,KM0}.
The main new step taken in \cite{KM0,MeissnerNicolai2019} consisted in extending these
considerations to the eight massive gravitinos which split as
\beq\label{GravCharges}
         \left({\bf 3}\,,\,\frac13\right) \oplus \left(\bar{\bf 3}\,,\,-\frac13\right)
         \oplus \left({\bf 1}\,,\,\frac23\right) \oplus \left({\bf 1}\,,\, -\frac23\right)
\eeq
under SU(3)$\,\times\,$U(1)$_{em}$. All gravitinos would thus carry 
{\em fractional electric charges}. If one identifies the SU(3) in (\ref{GravCharges}) 
with SU(3)$_c$ as in \cite{MeissnerNicolai2019}, a complex triplet of gravitinos would 
be subject to strong interactions. Importantly, the U(1)$_{em}$ charge 
assignments for the gravitinos include the spurion shift needed for matching 
the spin-$\frac12$ sectors \cite{MeissnerNicolai2019}.
The explanation of how to extend the spurion shift  to the gravitinos
requires a detour via $D=11$ supergravity and the enlargement of the 
SU(8) R-symmetry of $N\!=\!8$ supergravity to $K({\rm E}_{10})$ \cite{KN}.
Although the combined spin-$\frac12$ and 
spin-$\frac32$ content would thus coincide with the fermionic part of 
the $N\!=\!8$ supergravity multiplet, the underlying theory would need to be 
a very specific, but as yet unknown, extension
of (gauged) $N\!=\!8$ supergravity, which in its original form \cite{CJ,dWN} 
cannot be correct for reasons that have been known for more than 40 years.
Such a theory would almost certainly require a framework beyond space-time
based quantum field theory, but most likely different from conventional string theory.

In this note we would like to follow up on this line of thought and further
explore options towards linking $N\!=\!8$ supergravity to real physics, beyond 
matching its spin-$\frac12$ fermion content with that of the SM. More specifically, 
we would like to search for possible manifestations and experimental signatures 
in the spin-$\frac32$ sector which could support the above scenario. In comparison
with currently popular proposals for extending the SM, which usually come with a lot of 
extra new particles (especially in the context of low energy supersymmetry), 
the present scheme is thus very minimalistic. Nevertheless we believe that such 
a novel approach, also to the DM problem, is amply justified in view of strong indications 
that the SM could simply survive more or less {\em as is} all the way up to the 
Planck scale, and in view of decades of failed attempts to discover any signs 
of new physics beyond the SM with more conventional approaches. As we will 
argue here, the MACRO experiment \cite{MACRO0,MACRO}, which was 
originally set up for a very different purpose,  may offer some interesting new 
perspectives in this direction that merit further investigation.

\section{Superheavy gravitinos as dark matter candidates}
 
An important consequence of (\ref{GravCharges}) is that, due to their 
fractional charges the gravitinos cannot decay into SM fermions, and 
are therefore stable independently of their mass. Their stability against 
decays makes them natural candidates for DM \cite{MeissnerNicolai2019} 
(let us note that exotic fractionally charged stable states can also appear 
in certain supersymmetric GUT-type string compactifications \cite{Far}).
While a very large mass is strongly suggested  by the absence 
of low energy supersymmetry at LHC, an equally compelling 
argument for large mass are the bounds on the charge $q$ of any putative 
DM particle of mass $m_X$ derived in \cite{milli1,milli2,NNP}.
These bounds were derived with a perspective different from the present 
one, namely bounding the charge of `low mass' (in comparison with the 
Planck scale) DM candidates, by considering various astrophysical
constraints (structure formation and CMB, relic density, DM halo constraints,
{\em etc.}); for instance, \cite{milli1} considers only  DM candidates of 
mass $m_X \lesssim 10^4$ GeV, for which $q < 10^{-4}$. Taking the opposite 
view of this analysis as providing a {\em lower} bound on the mass of supermassive 
particles with given $\cO(1)$ charges therefore constitutes an extrapolation
well beyond the mass range considered in these references, and therefore
the inequalities derived there must be taken with a grain of salt.
Nevertheless, extrapolating the plots in figure 1 of \cite{milli1} to larger mass values
one sees that the bounds are actually much less constraining for masses close 
to the Planck scale. It then follows that gravitinos with charges (\ref{GravCharges}) 
can be comfortably within the allowed range to be viable DM candidates if 
their mass is close to, but still 
below, the Planck scale (if the Schwarzschild radius of
the particle were to exceed its Compton wave length, we would be dealing
not with a stable particle, but with a mini-black hole which would immediately 
decay by Hawking radiation).  As we argued in \cite{MeissnerNicolai2019} the
strongly interacting gravitinos would have mostly disappeared during the cosmic
evolution (but could play a role in explaining ultra-high energy cosmic rays
and the predominance 
of heavy ions in such events \cite{MN3}). By contrast, the abundance of the 
color singlet gravitinos cannot be estimated since they were never in thermal 
equilibrium, but one can plausibly assume their abundance in first approximation
to be given by the average DM density inside galaxies \cite{WdB}, {\it viz.}
\beq\label{DM}
\rho_{DM} \,\lesssim \,0.3\cdot 10^6 \GeV\cd m^{-3}  
\eeq 
(the average in the Universe is a million times smaller). 
If DM were entirely made out of nearly Planck mass particles with an assumed 
mass of $10^{18}$ GeV, this would amount to  $\sim 3\cdot 10^{-13}$ 
particles per cubic meter within galaxies Furthermore assuming an 
average velocity of 30 km$\,\cdot\,$s$^{-1}$ this yields the flux estimate
\beq\label{flux}
\Phi \;\sim \; 0.03 \, {\rm m}^{-2} {\rm yr}^{-1} {\rm sr}^{-1}
\eeq
We emphasize that, in addition to the unknown mass of the particle, 
there remain important uncertainties which could shift this estimate  in
either direction. One concerns possible inhomogeneities in the DM distribution 
within galaxies or stellar systems, which could lead to either a local 
depletion or to a local enhancement of the DM density (\ref{DM}) 
in the vicinity of the earth (see {\em e.g.} \cite{DK}
for an early discussion of this point). The other is, of course, the average 
velocity of superheavy DM particles w.r.t. the earth (which is $\sim 30$ km$\,\cdot\,$s$^{-1}$ 
if the DM particles are bound to the Solar System, and $\sim 300$ km$\,\cdot\,$s$^{-1}$ 
if they are bound to our Galaxy). In the end this will be a matter for observation to decide.

A distinctive feature of the present proposal is that the DM gravitinos {\em do} 
participate in SM interactions with couplings of order $\cO(1)$. In this sense,
our DM candidates are not dark at all, but simply too faint to be seen
directly because of their extremely low abundance! This is in contrast 
to other scenarios involving supermassive DM particles which are assumed 
to have only (super-)weak and gravitational interactions with SM matter
\cite{RGS,BJRR,CK,Markov,ACN,Sriv,SHDM1,SHDM2,PIDM,PIDM1,SHDM6}, 
and which are mainly motivated by inflationary cosmology, whence
the mass of those DM constituents would still be well below the Planck 
scale, on the order of the scale of inflation $\lesssim 10^{16}\,$ GeV.
By contrast the gravitinos in (\ref{GravCharges})  could in principle be detected 
if a way could be found to overcome their low abundance. 
A superheavy electrically charged particle could easily pass 
through the earth without deflection, leaving  
a very straight but tiny ionized track in the earth's crust.

This leaves us basically with two options for discovery. 
Either one searches for traces of such tracks in old and very
 stable rock with a paleodetector, or otherwise one sets up an underground 
detector with sufficiently large fiducial area/volume and waits for the candidate  
particle to come by. The paleodetector option has been tried in the past with 
MICA samples \cite{MICA0,MICA}, again to search for magnetic
monopoles;  a general difficulty here is that the 
tracks would have to remain unaffected by geological processes 
over very long times, and the detection technique must be such as not 
to destroy the tracks (this favors MICA which comes with a naturally
layered structure). The other and perhaps more promising option is to look for ionized
tracks with suitable underground detectors and time of flight measurements, 
focusing on {\em slow} ionizing particles.
The main background would come from cosmic ray muons, but a possible
way to rule those out would be to look for slow particles {\em moving bottom up}
which must have traversed a substantial part of the earth before being
registered by the detector. In this context, the possible relevance of 
the MACRO experiment \cite{MACRO0,MACRO} with its very large exposure 
time and surface area was already pointed out  in \cite{MeissnerNicolai2019}, 
where it was suggested to have a second look at the data collected  over 
many years. This is what we will now do, focusing on one special event.

\section{The MACRO experiment and a special event}

The MACRO experiment \cite{MACRO0,MACRO} was originally designed to search for magnetic
monopoles, finishing with a null result after several years of taking data. 
We refer to the summary paper \cite{MACRO} for a detailed description of the 
experiment and of what the detector was capable of doing, as well as a summary
of the collected results. The search covered a large part of parameter space, including 
the full range of velocities from relativistic particles down to `slow' particles 
with $\beta \sim 4 \times 10^{-5}$, coming in from all directions. In this way the 
detector was able to search not only for magnetic monopoles, but also for other, 
and unknown kinds of ionizing particles, including fractionally charged particles. 
Results of the latter search which concentrated specifically on {\em lightly
ionizing particles} (LIPs) appeared in a separate publication \cite{MACRO1}.

While \cite{MACRO} mentions 40 events (out of a total of about 35\,000) that were 
subsequently discarded as spurious and not further discussed,
Ref. \cite{MACRO1} reports one special event of a type different from an 
expected monopole signal. The relevant information about  this event is contained 
in figure 3 of \cite{MACRO1} which we here reproduce for the reader's 
convenience, together with its figure caption. 
\begin{figure}
\centering
\null\hfill\includegraphics[width=.8\textwidth, clip=true, trim = 115mm 180mm 0mm 20mm]{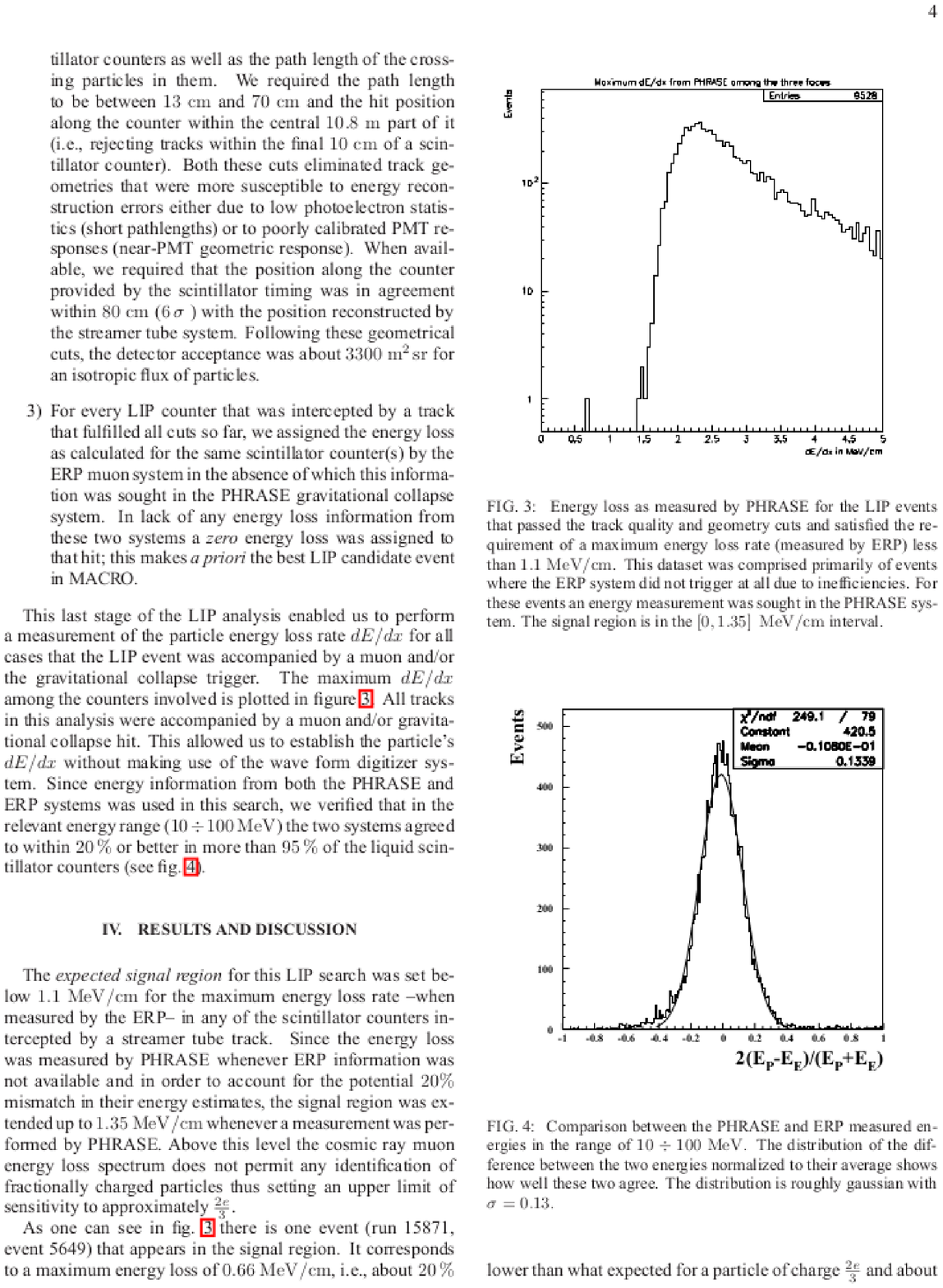}
\caption{Energy loss [...]
for the LIP events
that passed the track quality and geometry cuts and satisfied the requirement
of a maximum energy loss rate. 
[...]
The signal region is in the [0, 1.35] MeV/cm interval. Figure copied from \cite{MACRO1}.}
\end{figure}
Let us also quote excerpts from the accompanying 
part of the text in \cite{MACRO1} which says:
``As one can see [...] there is one event (run 15871,
event 5649) that appears in the signal region. It corresponds
to a maximum energy loss of 0.66 MeV/cm, i.e., about 20\% lower 
than expected for a particle of charge $2e/3$ and about
a factor of 3 higher than what expected for a particle of charge
$e/3$ . Three scintillator counters were involved in this trigger; the
first in one of the upper vertical layers, the second in the central
horizontal layer and the third in the lower horizontal layer.
[...] The position along the counter for this particular
box measured by the PHRASE and by the streamer
tube track geometry were in agreement (within 15 cm). We
have examined this event by hand relying primarily on the
wave forms as recorded for all the counters involved in the
trigger. The apparent amplitude of the recorded wave forms
was consistent with the energy thresholds [...].
Having three scintillator counters involved in the
trigger we have checked for a consistency in the relative timing
of them with the crossing of a single particle of constant
velocity. The relative timing between the counter in the upper
part of the detector and that in the central part was consistent
with the passage of a relativistic particle coming from above
while the relative timing between the box in the lower part of
the detector and any of the other two hits was consistent with a
slowly moving upward-going particle. We thus discarded this
event from the signal region.'' 

The main reason for discarding this event was therefore not some obvious
instrumental malfunction, but rather the fact that there appears 
to be no way to reconcile all three scintillator signals (as well 
as the fact that the signal does not conform with expectations for 
a magnetic monopole \cite{MACRO0,MACRO}), whence one
concludes that one of the three signals must be ascribed to a different origin.
A crucial additional fact is that there is unmistakable and independent evidence 
from the streamer tubes for a single particle track running through the whole detector.
The `obvious' interpretation of this event would seem to be in terms of
a relativistic particle corresponding to a  cosmic ray muon 
coming from above, and this remains perhaps the more plausible explanation.
However, this interpretation would not only require disregarding the 
earlier scintillator signal in the lower part of the detector, but also 
explaining why a third signal in the same scintillator, coincident with the 
central and upper scintillators, is missing. It would furthermore require an 
explanation why the energy loss rate is well below the allowed minimum in Fig.~1
corresponding to the minimum ionization point of the Bethe-Bloch curve \cite{PDG}
(which is reached for $\beta\gamma \sim 5$). A charge $|q| = \frac23$ (or less) 
is simply not compatible with a muon.

By contrast, the alternative second interpretation with a slow particle moving upward
would require only the signal in the upper scintillator to be due to some other cause. 
While it appears that the question cannot be finally resolved on the basis of 
the existing MACRO data, we here wish to raise attention to the possibility
that the signal in the upper layer could have masked the presumed later 
arrival of the slow particle in the upper detector. As we argue below, a slow 
particle would not have been registered at all without such an 
accompanying signal, on account of the trigger vetoes imposed for the LIPs search!
The event could then correspond to the passage of a `slow' charged particle
through the MACRO detector. To see why such a particle would be only lightly
ionizing we recall that the ionization rate increases with decreasing $\beta$ 
like $\propto \beta^{-2}$, but then drops again
rapidly for $\beta < 0.01$ (Lindhard-Scharff regime) \cite{PDG}. 
Even without a full fledged quantum mechanical calculation,  and
perhaps oversimplifying things a bit, this
can be understood by means of a very simple classical argument. For a head-on 
collision of a supermassive particle of speed $\beta$ with an electron,
the maximum velocity change imparted to the electron would be $2\beta$,
whence the maximum energy carried away by the electron would be 
$2m_e\beta^2 \sim 1 \, {\rm MeV} \times \beta^2$. For $\beta \lesssim 10^{-3}$
this is much less than the minimum ionization energy required to free the electron
from the atom (a proper quantum mechanical treatment, which is currently
not available, would need to take into account the interaction of the supermassive
particle with the whole cloud of electrons). Besides, for this range of 
$\beta$ the loss of energy by elastic collisions gets  more and more 
important with decreasing $\beta$. However, for a slow positively charged 
supermassive particle another process could become relevant: due to its large 
mass its motion is not affected at all by the atom, while thanks to its slow motion 
it can spend enough time in the electron cloud to `drag along' an electron, forming 
a lightly bound fractionally charged state that moves along with unchanged speed. 
Having removed the electron from the atom, it thus leaves 
behind an ion that can be detected by a streamer tube or a drift chamber,
but not a scintillator. This might explain why all streamer tubes registered
the track, but not all scintillators produced accompanying signals for the
event in question. Otherwise, in this regime, the energy losses would 
be dominated by nuclear recoils (which is not what MACRO was looking for),
which incidentally offers another possible avenue for testing our hypothesis.
Furthermore, in the absence of more detailed calculations, the charge 
of the particle can  no longer be straightforwardly deduced. 

If we therefore proceed with  this
assumption we have two additional indications for the correctness
of our hypothesis, as emphasized in the above quote from \cite{MACRO1}, namely
\begin{itemize}
\item the track as seen by the streamer tubes was 
         consistent with the tracks in the lower and central scintillators; and
\item the time of flight was consistent with a slow particle moving 
         bottom up in the lower and central scintillators 
\end{itemize}
With these additional consistency checks let us re-iterate that a `slow' particle 
moving bottom up would be difficult to explain in terms of known physics. First
of all, going up, it obviously cannot be a muon. Second, it cannot be a magnetic 
monopole, nor a dyon, because monopoles generally are not expected 
to be lightly ionizing because of their large magnetic charge, although they 
may in principle be able to traverse the  earth \cite{MM1}. This is also the
energy range where the relation between the energy loss and the 
light yield (which is what is measured by the photomultipliers) 
is no longer linear \cite{MM2}. The same comment applies to dyons
for  which the energy loss would be even bigger
(in principle an electrically neutral monopole can acquire a very 
small electric charge proportional to the CP violating $\theta$-angle by means 
of the Witten effect \cite{Witten}; however, given the known upper limits on the
value of $\theta$ the ionization would be dominated by magnetic interactions).
Third, the full track as reconstructed from the streamer tubes
cannot be the result of a radioactive decay in the surrounding rock, 
since such products have energies of at most several MeV, so they could not 
penetrate the scintillator more deeply than a few centimeters. Therefore a
superheavy fractionally charged particle seems to be the most plausible 
explanation if one adopts our hypothesis.
 
While \cite{MACRO1} does not explicitly quantify what `slow' means 
a more precise knowledge of the velocity would be useful as it would enable 
us to make a first estimate of the expected gravitino flux. If the MACRO detector
had been set up to search for slow particles such as a supermassive gravitino,
the expected event rate according to the estimate (\ref{flux}) would amount
to several events per year. 
However, the trigger used in \cite{MACRO1} was sensitive only to {\em fast} 
lightly ionizing particles. As stated in \cite{MACRO1}: ``Particles which
have a velocity lower than 0.25$c$ are not guaranteed to pass
through the detector quickly enough to insure that the LIP
trigger will detect a coincidence in the faces of the scintillator system''. Hence 
without an accompanying coincident triggering signal any slower particle 
would have escaped detection, and thus possible events involving {\em only} a slow 
particle could have been missed. Therefore the rate of one event for the five-year 
cycle covered by \cite{MACRO1} could be a significant underestimate of the 
actual abundance and flux rate for supermassive gravitinos. Conversely, the trigger 
settings used for the original monopole search \cite{MACRO0,MACRO} which did
include velocity ranges down to $\beta \sim 4\cdot 10^{-5}$, did not allow for lightly 
ionizing particles, thus explaining why no further events of this type were observed, 
even for a flux close to the estimated value (\ref{flux}). For this reason
we cannot at this point reliably deduce the actual gravitino flux in the vicinity 
of the Earth from the given data. This, as well as the confirmation (or refutation)
of our hypothesis, would require a dedicated new experiment, concentrating on
slow and lightly ionizing particles.

\section{Conclusions}
Re-inspection of the special MACRO event reported in \cite{MACRO1} 
has revealed the possibility of an explanation different from the `obvious' one 
in terms of a cosmic ray muon, namely

\begin{itemize}
\item a slow particle {\em moving bottom up}, which thus must have
         traversed a substantial part of the earth; and
\item partial evidence that this particle carries fractional charge. 
\end{itemize}
Nevertheless, in order to find out whether this outlier event is real physics 
or just a fluke, and to proceed from circumstantial evidence to an actual discovery,
further and independent confirmation is obviously needed, 
and we hope that the present paper can motivate a
dedicated effort that will decide the issue  in the not-so-distant future.

On the other hand, corroboration of our new interpretation 
of this event would have dramatic implications. In particular, it would 
bring $N\!=\!8$ supergravity back into focus
for unification, although in an unexpected way. We emphasize that the 
considerations leading to (\ref{GravCharges}) are so far purely kinematical,
and that the dynamics underlying the present scheme remains unknown,
possibly requiring a framework beyond space-time based quantum field theory.
Nevertheless our findings may indicate that $N\!=\!8$ supergravity could be closer 
to the truth than is widely thought (as is also suggested by the finiteness properties
of the theory \cite{Bern} and various anomaly cancellations
\cite{Marcus,Kallosh,Zvi,MN4}). We also note that the spurion shift required 
to match the spin-$\frac12$ sectors of the theory with three generations
of quarks and leptons \cite{GM},  and here extended to the gravitinos \cite{KM0},
appears to be  incompatible with space-time supersymmetry. This could mean 
that, contrary to many expectations, (maximal) space-time supersymmetry might 
not be a relevant concept for unification after all, but, through its fermionic 
(spin-$\frac12$ and spin-$\frac32$) content, merely a
theoretical crutch to guide us to the right answer.

\vspace{0.5cm}
\noindent
 {\bf Acknowledgments:} 
 We would like to thank Barry Barish for helpful correspondence. Furthermore, 
 we are greatly indebted to Erik Katsavounidis for alerting us to \cite{MACRO1}
 and the strange outlier event, and for many explanations and clarifications 
 concerning the MACRO experiment without which this article would not have 
 seen the light of the day. K.A.~Meissner was partially 
 supported by the Polish National Science Center grant UMO-2020/39/B/ST2/01279.
 The work of  H.~Nicolai has received funding from the European Research 
 Council (ERC) under the  European Union's Horizon 2020 research and 
 innovation programme (grant agreement No 740209).

\end{document}